\documentclass[floatfix,nofootinbib,twocolumn,preprintnumbers,superscriptaddress,notitlepage]{revtex4-1}

\usepackage{times}
\usepackage{xcolor}
\usepackage{amsfonts}
\usepackage{amsmath}
\usepackage{amssymb}
\usepackage{bm}
\usepackage{dcolumn}
\usepackage{enumerate}
\usepackage{epsfig}
\usepackage{graphicx}
\usepackage{graphics}
\usepackage[latin1]{inputenc}
\usepackage{latexsym}
\usepackage{rotating}
\usepackage{hyperref}
\usepackage[caption=false]{subfig}
\usepackage[normalem]{ulem}
\usepackage{tabularx}

\newcommand{\<}{\begin{equation}}
\newcommand{\?}{\end{equation}}

\newcommand{\cX}{\mathcal{X}}
\newcommand{\cY}{\mathcal{Y}}
\newcommand{\cZ}{\mathcal{Z}}

\DeclareMathOperator{\atantwo}{arctan2}

\begin{document}

\title{Distinguishing binary black hole precessional morphologies with gravitational wave observations}

\author{Nathan~K.~Johnson-McDaniel}
\affiliation{Department of Physics and Astronomy, The University of Mississippi, University, Mississippi 38677, USA}
\author{Khun Sang Phukon}
\affiliation{Nikhef - National Institute for Subatomic Physics, Science Park, 1098 XG Amsterdam, The Netherlands}
\affiliation{Institute for High-Energy Physics, University of Amsterdam, Science Park, 1098 XG Amsterdam, The Netherlands}
\affiliation{Institute for Gravitational and Subatomic Physics, Utrecht University, Princetonplein 1, 3584 CC Utrecht, The Netherlands}
\affiliation{School of Physics and Astronomy and Institute for Gravitational Wave Astronomy,\\University of Birmingham, Edgbaston, Birmingham, B15 9TT, United Kingdom}
\author{N.~V.~Krishnendu}
\affiliation{Max Planck Institute for Gravitational Physics (Albert Einstein Institute), Callinstr.~38, D-30167 Hannover, Germany}
\affiliation{Leibniz Universit{\"a}t Hannover, D-30167 Hannover, Germany}
\affiliation{International Centre for Theoretical Sciences (ICTS), Survey No. 151, Shivakote, Hesaraghatta, Uttarahalli Hobli, Bengaluru, 560089}
\author{Anuradha Gupta}
\affiliation{Department of Physics and Astronomy, The University of Mississippi, University, Mississippi 38677, USA}

\date{\today}

\begin{abstract}
The precessional motion of binary black holes can be classified into one of three morphologies, 
based on the evolution of the angle between the components of the spins in the orbital plane: 
Circulating, librating around 0, and librating around $\pi$. These different morphologies can be 
related to the binary's formation channel and are imprinted in the binary's gravitational wave signal. 
In this paper, we develop a Bayesian model selection method to determine the preferred spin morphology of a detected binary black hole.
The method involves a fast calculation of the morphology which allows us to restrict to a specific morphology in the Bayesian stochastic sampling. 
We investigate the prospects for distinguishing between the different morphologies using gravitational waves 
in the Advanced LIGO/Advanced Virgo network with their plus-era sensitivities. 
For this, we consider fiducial high- and low-mass binaries having different spin magnitudes and signal-to-noise ratios (SNRs).
We find that in the cases with high spin and high SNR, the true morphology is strongly favored with $\log_{10}$ Bayes factors $\gtrsim 4$ compared to both
alternative morphologies when the binary's parameters are not close to the boundary between morphologies. However, when the binary parameters are close to the boundary between morphologies, only one alternative morphology is strongly disfavored.
In the low-spin, high-SNR cases, the true morphology is still favored
with a $\log_{10}$ Bayes factor $\sim 2$ compared to one alternative morphology, while in the low-SNR cases the $\log_{10}$ Bayes factors are at most $\sim 1$ for many binaries. We also consider the gravitational wave signal from
GW200129\_065458 that has some evidence for precession (modulo data quality issues) and find that there is no preference for a specific morphology. Our method for restricting the prior to a given morphology is publicly available through an easy-to-use Python
package called \verb|bbh_spin_morphology_prior|.
\end{abstract}

\maketitle

\section{Introduction}
\label{sec:intro}
The LIGO-Virgo observing runs during 2015-2020 have discovered $\sim100$ binary coalescence events, mostly consisting of binary black hole (BBH) mergers (see \cite{LIGOScientific:2021djp, Nitz:2021zwj, Olsen:2022pin} for the latest results).
With these many detections, we are now able to constrain present-day binary merger rates and their evolution with redshift as well as the mass and spin distribution of the components of stellar-mass BBHs in the Universe~\cite{LIGOScientific:2021psn}. These discoveries thus help us study the different pathways for compact object binary formation and associated physical processes. 
In the coming years, the number of observed BBHs will rapidly increase with inclusion of the KAGRA~\cite{KAGRA:2020tym} and planned LIGO-India~\cite{LigoIndia2011} detectors in the network, as well as upgrades~\cite{Aasi:2013wya} to the existing Advanced LIGO~\cite{LIGOScientific:2014pky} and Advanced Virgo~\cite{VIRGO:2014yos} detectors.

Understanding astrophysical processes involved in BBH formation has a particular importance in gravitational wave (GW) astronomy. Many potential astrophysical processes can lead to the formation of the observed stellar-mass BBHs by ground-based GW detectors. These formation scenarios broadly fall into two categories: the field channel in which BBHs are formed from isolated evolution of massive stars in binaries and the dynamical channel in which BBHs are created through dynamical encounters in dense environments such as globular clusters or disks of active galactic nuclei (for reviews, see \cite{Mapelli:2021taw, Mandel:2018hfr}). Apart from these astrophysical processes, primordial black holes formed in the early Universe may also form binaries dynamically and contribute to the observed population (see, e.g.,~\cite{Bird:2016dcv,Carr:2019kxo}). The formation mechanisms leave intriguing features in the distribution of intrinsic parameters of binaries.

Black hole spin orientations are interesting 
parameters that strongly depend on the formation scenarios of BBHs. Black holes in field binaries are expected to have spins preferentially aligned with the orbital angular momentum, whereas spins have no preferred orientation in binaries formed through the dynamical channel~\cite{2000ApJ...541..319K, Rodriguez:2016vmx}. In the post-Newtonian (PN) picture, the spin-orbit misalignment induces relativistic precession of the spins and orbital angular momentum about the binary's total angular momentum as the binary evolves (e.g.,~\cite{Apostolatos:1994mx, Kidder:1995zr, Racine:2008qv, Schnittman:2004vq}). The precessional effects are encoded into the emitted GW signal through characteristic amplitude and phase modulations. These rich characteristics in GW signals enable better measurement of parameters of the source such as masses and spin components in precessing systems~\cite{vanderSluys:2007st, vanderSluys:2008qx, Farr:2014qka, Vitale:2014mka,Pratten:2020igi,Biscoveanu:2021nvg,Knee:2021noc,Krishnendu:2021cyi}. However, these characteristics make search and parameter estimation on such signals challenging due to the large dimensionality of the parameter space and correlations between various parameters~\cite{Harry:2016ijz,Veitch:2014wba}.

\begin{figure}
\includegraphics[width=0.5\textwidth]{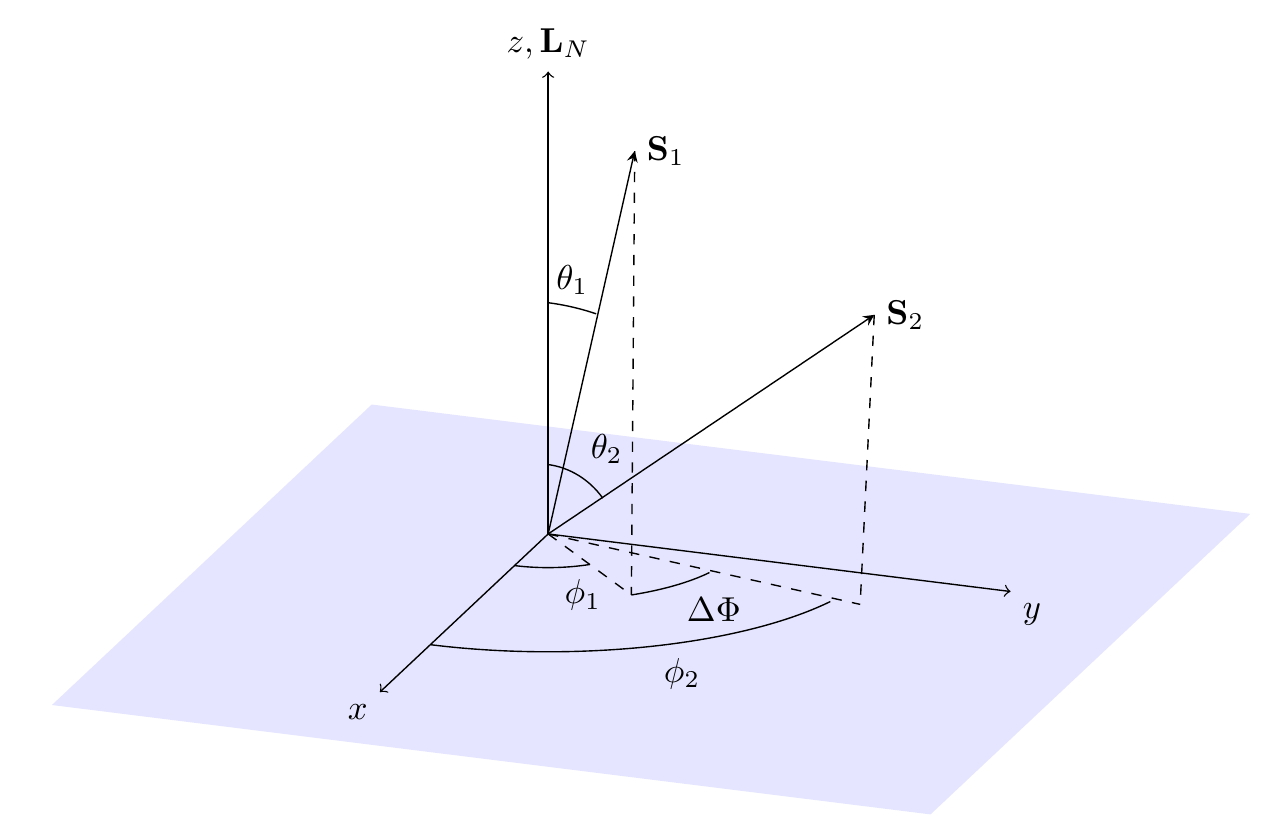}
\caption{\label{fig:angles} The angular momentum vectors of a precessing BBH in a reference frame whose $xy$-plane is the orbital plane of the binary, with the $z$-axis along the Newtonian orbital angular momentum vector $\mathbf{L}_N$. The angles $\theta_1$ and $\theta_2$ are the tilt angles of spins of two components of the binary. The orbital plane angles of the two spins are denoted by $\phi_1$ and $\phi_2$, and $\Delta \Phi = \phi_2 - \phi_1$ is the difference between these in-plane angles.}
\end{figure}

One of the exciting features of the spin-precession dynamics is that precessing BBHs can reside in different spin morphologies during their evolution. These spin morphologies are generalizations of the spin-orbit resonances first identified by Schnittman~\cite{Schnittman:2004vq} and can be elegantly described in the PN approximation using the effective potential formalism given in~\cite{Kesden:2014sla,Gerosa:2015tea}. This formalism is also used to study spin morphologies of precessing binaries  in the equal mass limit~\cite{Gerosa:2016aus} and with non-zero eccentricity~\cite{Phukon:2019gfh}. In this formalism, the binary's spin dynamics can be classified into three spin morphologies based on the evolution of the angle $\Delta \Phi$ between the projections of two black hole spin angular momenta onto the orbital plane (see Fig.~\ref{fig:angles}), commonly referred to as $\phi_{12}$ in GW data analysis (e.g.,~\cite{Farr:2014qka,LIGOScientific:2021djp}). Over a precessional period, $\Delta \Phi$ can either circulate in the full range $[-\pi, \pi]$ or librate around either $0$ or $\pi$, never reaching the other. These correspond to the C, $\mathrm{L}0$, and $\mathrm{L}\pi$ morphologies, respectively. The librating morphologies reduce to Schnittman's spin-orbit resonances~\cite{Schnittman:2004vq} in the limit in which the libration amplitude reduces to zero.

The spin morphology is a property of the conservative dynamics and can change on the radiation reaction timescale. Binaries are in the C morphology at infinite separation, and can evolve to a librating morphology close to merger. It is possible to have multiple morphology transitions, where the binary evolves back to the C morphology after evolving into a librating morphology~\cite{Gerosa:2015tea}. However, in most cases, there is a single morphology transition. The evolution of morphologies is much slower than the evolution of spin tilts which happens on the precessional timescale. In \cite{Gerosa:2015tea}, it is shown that morphologies of a binary at a small orbital separation can be related to the binary's spin tilt angles at asymptotically large separation. In particular, one finds that populations of binaries that formed with different spin tilts evolve into sub-populations with distinctive $\Delta \Phi$ distributions at the small orbital separations to which ground-based GW detectors are sensitive~\cite{Schnittman:2004vq, Kesden:2010yp, Kesden:2010ji,Berti:2012zp, Gerosa:2013laa}.
Spin morphology quantifies the evolution of  $\Delta \Phi$ or precession dynamics in general in a precession cycle.  Thus, spin morphologies of BBHs in the band of ground-based GW detectors are indicative of the spin configurations of binaries at formation~\cite{Gerosa:2015tea}. In particular, morphology measurements using GWs enable us to constrain certain formation channels or different evolutionary scenarios of stellar progenitors of field BBHs in some cases, e.g., when tidal alignment and natal kicks are significant \cite{Gerosa:2013laa,Gerosa:2018wbw}. See also~\cite{Steinle:2022rhj} for more recent work relating the morphologies in the ground-based detector band with astrophysical processes in the binaries' formation.

Recognizing the astrophysical importance of spin morphologies, many studies have investigated issues of detection and characterization of BBHs in resonant configurations (e.g., \cite{Afle:2018slw}) or $\Delta \Phi$ measurement. Initial studies with PN waveforms found that $\Delta \Phi$ is generally a poorly constrained quantity in the band of ground-based GW detectors~\cite{Vitale:2014mka}, though it may be possible to constrain it for highly spinning binaries with suitable inclination angles~\cite{Trifiro:2015zda}. However, more recent studies show that one may get reasonably good constraints on $\Delta \Phi$ at a reference frequency close to merger from loud signals~\cite{Varma:2021csh} when using full inspiral-merger-ringdown waveform models, including the IMRPhenomXPHM~\cite{Pratten:2020ceb} and NRSur7dq4~\cite{Varma:2019csw} models that we use here.
Previously, \cite{Gupta:2013mea} presented a framework to compute time-domain PN inspiralling templates for BBHs in spin-orbit resonances and showed that a BBH in a spin-orbit resonance can be distinguished from other binaries. Ref.~\cite{Gerosa:2014kta} examined PN waveforms from resonant binaries and assessed their distinguishability, while Ref.~\cite{Trifiro:2015zda} performed parameter estimation with those waveforms, focusing on parameters which are important for resonances, and inferred the probability for the binary to be in a given morphology by counting the number of posterior samples in each morphology. These studies considered binaries with total mass less than $35M_\odot$ due to the use of a PN waveform model. 

Recently, Varma {\it et al.}~\cite{Varma:2021xbh} presented constraints on the distribution of in-plane spin angles and $\Delta \Phi$ in the population of high-mass GWTC-2 events using a hierarchical Bayesian framework and found weak evidence for resonant configurations in the BBH population. Their analysis inferred these angles at a fixed dimensionless time soon before merger, which they found to give better constraints on the individual angles $\phi_{1,2}$ than inferring the angles at a fixed dimensionful reference frequency, as is commonly done in GW data analysis, though the constraints on $\Delta\Phi$ are not significantly improved. Specifically, they picked the reference time of $100M$ before merger, where $M$ is the binary's total mass. In~\cite{Varma:2021csh}, they also considered using a dimensionless reference frequency (so scaling inversely with the binary's total mass) and found that this also gives an improvement in the constraints similar to that obtained with the dimensionless reference time. In this paper, we consider a dimensionful reference frequency, as is standard, and still find that it is possible to infer the morphology with high confidence in certain cases. We leave considering a dimensionless reference point for future work.

Additionally, Gangardt~\emph{et al.}~\cite{Gangardt:2022ltd} considered the constraints it is possible to place on the morphology of the BBHs in GWTC-3 by counting the number of posterior samples in each morphology, finding that there is no significant preference for any morphology. In particular, they find that the fractions of posterior samples in the librating morphologies agree with those in the prior after conditioning on the individual masses and effective spin (while the morphology depends sensitively on additional spin degrees of freedom, notably $\Delta\Phi$).

In this paper, we develop a Bayesian model selection method to distinguish the three possible spin morphologies in BBHs.  We demonstrate that at the detector sensitivities expected for the O5 observing run~\cite{Aasi:2013wya}, the true morphology of a BBH can be identified with very high confidence in optimistic cases such as high spins and high signal-to-noise ratios (SNRs). 
In our analysis, we consider comparable mass BBHs with different spin magnitudes, spin-orbit misalignments and SNRs and two total masses which are consistent with the over-dense regions of chirp  mass distribution inferred from GWTC-3~\cite{LIGOScientific:2021psn}. We also apply our method on the loud event GW200129\_065458 which is possibly a precessing system~\cite{LIGOScientific:2021djp,Hannam:2021pit,Payne:2022spz}, finding that no spin morphology is preferred. Our code to restrict the prior to a specific morphology in Bayesian stochastic sampling is publicly available as a Python package \verb|bbh_spin_morphology_prior|~\cite{morph_package}.

The remainder of the paper is structured as follows. In Sec.~\ref{sec:morph}, we discuss how to compute a given binary's morphology. In Sec.~\ref{sec:methods}, we give an overview of our Bayesian inference method and the simulated mock GW observations we consider to assess the performance of the method. Sec.~\ref{sec:results} contains the results from our analysis on simulated GW signals and GW200129\_065458. Finally, we conclude in Sec.~\ref{sec:concl}, highlighting the relevance of our findings to GW astrophysics. Throughout the paper, we  work in geometric units $(G=c=1)$.


\section{Computation of precessional morphologies}
\label{sec:morph}
We consider a BBH system in the PN approximation, with individual masses $m_{1,2}$, dimensionless spin vectors $\boldsymbol{\chi}_{1,2}$, and Newtonian orbital angular momentum vector $\mathbf{L}_N$. For our purposes, the tilt angles $\theta_{1,2}$ between $\mathbf{L}_N$ and $\boldsymbol{\chi}_{1,2}$ (or equivalently the dimensionful spins $\mathbf{S}_{1,2}$) and the angle between the in-plane spin components $\Delta\Phi$ (all illustrated in Fig.~\ref{fig:angles}) will be the most important parameters. The effective spin $\xi := (m_1\boldsymbol{\chi}_1 + m_2\boldsymbol{\chi}_2)\cdot\hat{\mathbf{L}}_N/(m_1 + m_2)$, where the circumflex denotes a unit vector, will also be important for our discussion. The effective spin is often denoted by $\chi_\text{eff}$ in the literature but here we follow the notation of Gerosa~\emph{et al.}~\cite{Gerosa:2015tea}.

As discussed in~\cite{Gerosa:2015tea}, there is a simple method for computing the morphology of a BBH at a given reference frequency using the $2$PN orbit-averaged equations. Here one is able to use the conservation of the effective spin $\xi$ to derive an effective potential for the evolution of the binary's total spin magnitude $S = \|\mathbf{S}_1 + \mathbf{S}_2\|$. One then obtains a cubic equation in $S^2$, whose two larger roots $S^2_\pm$ give the values of $S^2$ at the turning points of the binary's precessional evolution. One can thus determine the binary's morphology by evaluating $\Delta\Phi$ at each of $S^2_\pm$. For the C (circulating) morphology one obtains $\Delta\Phi = 0$ at one turning point and $\Delta\Phi = \pi$ at the other,\footnote{One can further distinguish between the common case where $\Delta\Phi = 0$ and $\pi$ correspond to $S_+^2$ and $S_-^2$, respectively, and the uncommon case where these correspond instead to $S_-^2$ and $S_+^2$ (see~\cite{Gerosa:2023xsx}, where these cases are denoted C$+$ and C$-$, respectively). However, we will not make this distinction in the current study.} while for the L$0$ and L$\pi$ morphology, one has the same value of $\Delta\Phi$ at both turning points ($0$ and $\pi$, respectively). We now describe our streamlined method for performing this calculation. This is released in an implementation~\cite{morph_package} that interfaces with the Bilby inference code~\cite{Ashton:2018jfp, Romero-Shaw:2020owr}.

We want to obtain the two larger roots of the cubic $S^6 + B S^4 + C S^2 + D = 0$, where the coefficients are given in Eqs.~(16) of~\cite{Gerosa:2015tea} and in a slightly different form in Appendix~B of~\cite{Chatziioannou:2017tdw}. Here we give them in the form we use in the code:
\<
\begin{split}
B &= -2 J^2 + 2 \xi L + \frac{(1 + q^2) L^2 - \cX}{q},\\
C &= (J^2 - L^2)^2 - 2\xi L (J^2 - L^2) + \frac{4q \xi^2 L^2}{(1 + q)^2}\\
&\quad + \frac{2(J^2\cX - L^2\cY)}{q} - \frac{2\xi L \cZ}{1 + q},\\
D &= \frac{L^2\cZ^2 - (J^2 - L^2)^2 \cX}{q} + \frac{2\xi  L (J^2 - L^2) \cZ}{1 + q},
\end{split}
\?
where $\cX := (1 - q)(q S_1^2 - S_2^2)$, $\cY := (1 - q)(S_1^2 - q S_2^2)$, and $\cZ := (1 - q)(S_1^2 - S_2^2)$. Here we are working in total mass = 1 units, $S_{1,2} = m_{1,2}^2\chi_{1,2}$ denote the magnitudes of the black holes' spin angular momenta, $q := m_2/m_1 \leq 1$ is the mass ratio, $J = \|\mathbf{L} + \mathbf{S}_1 + \mathbf{S}_2\|$ is the magnitude of the binary's total angular momentum, and $\mathbf{L}$ is its (PN) orbital angular momentum, with magnitude $L$.
We found that it is most efficient to solve the cubic analytically, using trigonometric functions, so there is no need for complex arithmetic. Specifically, we find that
\<
\begin{split}
S^2_- &= -\frac{B}{3} - \frac{2}{3}\sqrt{B^2 - 3C}\cos\left(\alpha + \frac{\pi}{3}\right),\\
S^2_+ &= -\frac{B}{3} + \frac{2}{3}\sqrt{B^2 - 3C}\cos\alpha,
\end{split}
\?
where $\alpha := \atantwo(3\sqrt{3\bar{\Delta}},-2B^3 + 9BC - 27D)/3$ and $\bar{\Delta} := B^2C^2 - 4B^3D + 18BCD - 4C^3 - 27D^2$ is (up to a constant factor) the discriminant of the equation.

We also simplify the expression for $\cos\Delta\Phi(S_\pm)$ used to obtain the morphology. Noting that all we need is the sign of this quantity, we start from Eqs.~(20) in~\cite{Gerosa:2015tea} and remove the $\sin\theta_A$ terms that have a single sign and multiply through by the magnitudes of the spins to obtain
\begin{widetext}
\<\label{eq:cosDeltaPhi_sgn}
\cos\Delta\Phi(S_\pm) \propto S_\pm^2 - S_1^2 - S_2^2 + \frac{2q}{(1-q)^2}\left(\frac{J^2 - L^2 - S_\pm^2}{2L} - \frac{q\xi}{1 + q}\right)\left(\frac{J^2 - L^2 - S_\pm^2}{2L} - \frac{\xi}{1 + q}\right),
\?
\end{widetext}
where the proportionality factor is positive. This expression is not well behaved for $q = 1$, due to the $1/(1-q)^2$ term, but this is not a concern for the present application, where one will never encounter exactly equal masses in the stochastic sampling. Since we only need to know the sign of $\cos\Delta\Phi(S_\pm)$, it is possible to use Descartes' rule of signs to determine the morphology without solving the cubic and thus without any need for trigonometric functions. However, we found that this was only a few percent faster than the more straightforward computation given above, so we kept the more straightforward version in our implementation.

Since this is only a $2$PN computation, we need to check that it still gives reliable results for the fairly relativistic binaries we are considering, with orbital velocities of $\sim 0.3$ at the reference frequency of $20$~Hz for the larger total mass ($75M_\odot$) we consider. To do this, we compare the results of the morphology computation and the $\Delta\Phi$ evolution given by the $3.5$PN point-particle, $3$PN spin order orbit-averaged evolution as implemented in the LALSuite~\cite{LALSuite} SpinTaylor evolution~\cite{SpinTaylor_TechNote} as well as the NRSur7dq4 numerical relativity (NR) surrogate model~\cite{Varma:2019csw} that does not use orbit-averaging. We do not just compare with the NRSur7dq4 model, since this is not available much earlier in the evolution than the reference frequency. The NRSur7dq4 spin dynamics are also given in terms of coordinate quantities from the NR simulation. Thus, they are not in the same coordinate system (and spin supplementary condition) as the PN equations. However, we expect these coordinate systems to agree well, given how the NR coordinate system is chosen, and find that this is indeed the case for the binaries we consider. In fact, \cite{Ossokine:2015vda} found that the PN spin evolution agrees well with the output of the NR simulations (computed using the same NR code that provides the waveforms used for the surrogate model). Moreover, the surrogate for the remnant properties (also described in~\cite{Varma:2019csw}) uses the SpinTaylor evolution to evolve the spins when the starting frequency is below the minimum available frequency in the surrogate model, and then uses the spins output by SpinTaylor as direct input to the surrogate evolution.

We find that one obtains good accuracy for the morphology calculation (comparing to the SpinTaylor evolution) when using the $1.5$PN expression for the orbital angular momentum in the $2$PN morphology expression instead of the Newtonian (i.e., $0$PN) expression for the orbital angular momentum that gives a consistent PN result (and is thus the one used in~\cite{Gerosa:2015tea}; we have checked that our calculation of the morphology agrees with that from the PRECESSION code~\cite{Gerosa:2016sys} when we use the Newtonian orbital angular momentum). We considered the orbital angular momentum expressions given up to $2.5$PN in Eq.~(4.7) of~\cite{Bohe:2012mr}. Here we orbit average the spins, as discussed in, e.g., Sec.~II of~\cite{SpinTaylor_TechNote} [which also gives the orbital angular momentum expressions in Eq.~(4)]. However, we find that the corrections to the orbital angular momentum above $1.5$PN have a small effect on the morphology. In fact, the $1.5$PN expression for the orbital angular momentum itself only gives a small correction to the $1$PN results in most cases. We thus use the $1.5$PN expressions for the orbital angular momentum (the first order at which there are spin contributions), since using higher-order expressions would lead to a more complicated and expensive morphology calculation. Note that the $n$PN spin order SpinTaylor evolution uses the $(n - 1.5)$PN order orbital angular momentum, and there is no $0.5$PN correction to the orbital angular momentum, so the $2$PN ($3$PN) evolution uses the $0$PN ($1.5$PN) orbital angular momentum.

\begin{figure}
\includegraphics[width=0.5\textwidth]{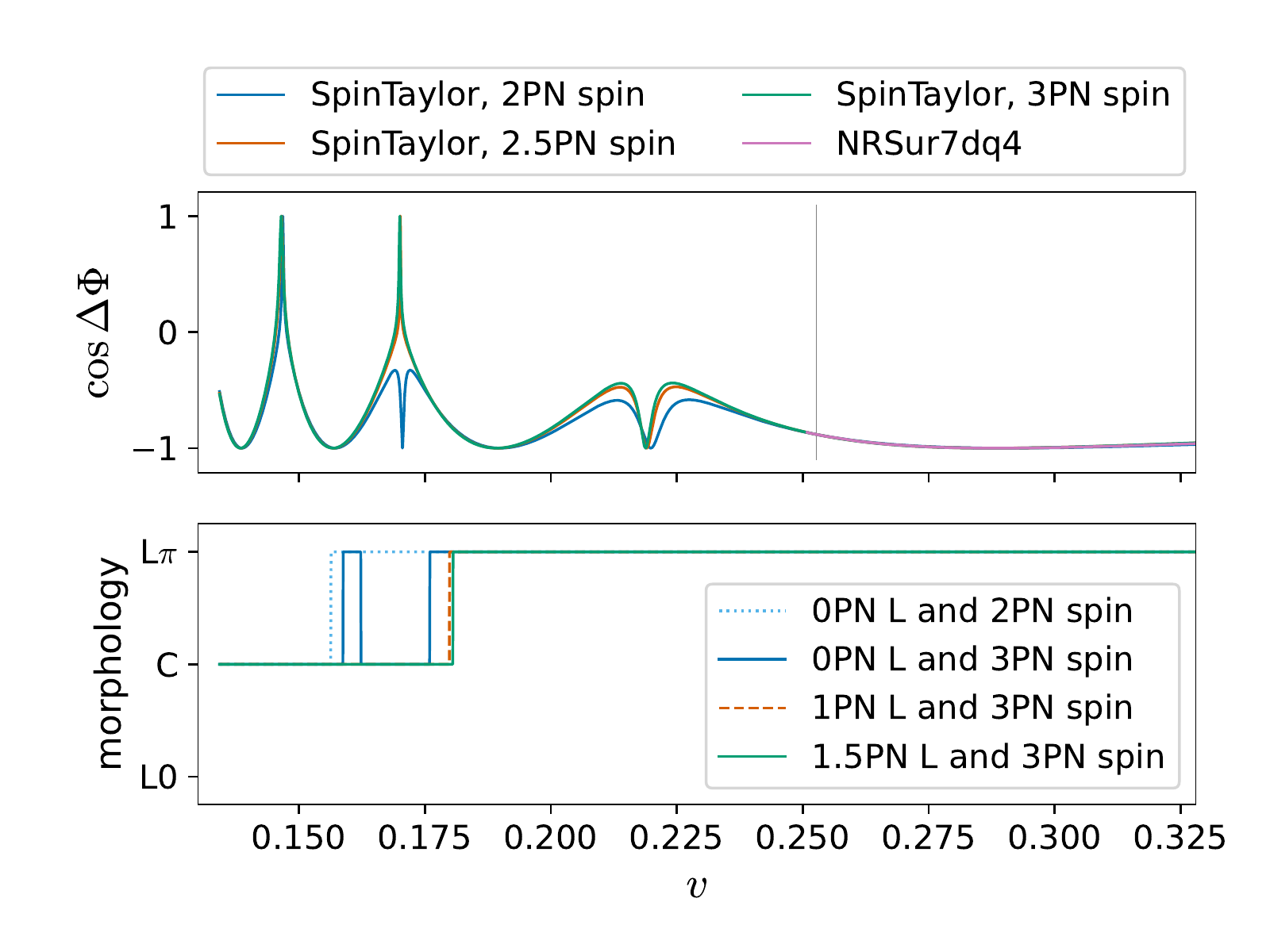}
\caption{\label{fig:morph_trans} The transition from the C to L$\pi$ morphology for an example binary (described in the text) versus the binary's orbital velocity $v$ as given by the SpinTaylor evolution (and different PN spin orders) and the NR surrogate evolution. The NR surrogate evolution is only available starting at $v \simeq 0.25$, shortly before the reference frequency ($20$~Hz, marked with a thin vertical line). The upper panel shows the evolution of $\cos\Delta\Phi$ for different spin order settings in the SpinTaylor evolution [with $n$PN spin order using the $(n-1.5)$PN orbital angular momentum], where the legend gives the spin order.
The lower panel shows the morphology prediction at each point of the evolution obtained using the spin angles from the $3$PN spin order evolution.
Here, we use different PN orders when computing the orbital angular momentum in the morphology calculation, illustrating the significant difference between the $0$PN case (solid blue line) and the higher PN orders (dashed orange and solid green lines). We also show the morphology prediction with the $0$PN orbital angular momentum applied to the spin angles from the $2$PN spin order evolution (dotted blue line in the lower panel) in order to illustrate that it agrees with the $2$PN
spin order evolution (solid blue line in the upper panel). We do not show the $2$PN and $2.5$PN orbital angular momentum results since they are almost indistinguishable from the $1.5$PN result.}
\end{figure}

In Fig.~\ref{fig:morph_trans}, we illustrate the transition from the C to L$\pi$ morphology for a case that gives a notable difference with the different PN orders. This is a specialized case, but in all cases using the $0$PN orbital angular momentum in the morphology calculation gives an earlier transition to the librating morphology than the higher-order terms do.\footnote{The fact that the $0$PN orbital angular momentum gives an earlier transition to the librating morphologies than the higher-order expressions do is to be expected: The $1$PN correction dominates and it always increases the orbital angular momentum, so one has to consider a larger orbital velocity to obtain the same magnitude of orbital angular momentum with the $1$PN expression as with the $0$PN expression.} However, in most cases there is no apparent transition back to the circulating morphology, as we find when applying the $0$PN orbital angular momentum morphology calculation to the results from the $3$PN evolution. The parameters used are a mass ratio of $1/1.22$, spin magnitudes of $\chi_1 = 0.708$, $\chi_2 = 0.843$, and spin angles (in radians) of $\theta_1 = 2.49$, $\theta_2 = 1.41$, and $\Delta\Phi = 2.65$. Here the angles are defined at an orbital velocity of $0.253$, which corresponds to a dominant mode GW frequency of $20$~Hz for the binary's total mass of $52.2M_\odot$. We use the SpinTaylorT4 approximant for all the results shown and found that the other two available approximants (SpinTaylorT1 and SpinTaylorT5) gave indistinguishable results in the region plotted in Fig.~\ref{fig:morph_trans}.

We find that the PN and surrogate evolutions agree well in all the cases we consider (parameters given in Table~\ref{table:angle_pars}), which gives us confidence that the orbit averaging used in the PN calculations is a good approximation for these purposes even relatively close to merger. This should be expected, since there are still tens of orbits in the final precessional cycle (which is the one during which the binary reaches a dominant GW frequency of $20$~Hz in the $75 M_\odot$ case).

The implementation of the morphology calculation for Bilby~\cite{morph_package} adds $\sim 2$~$\mu$s per sample to the time it takes Bilby to compute the prior on a $2.8$~GHz Intel Xeon E5-2680 processor, which is also $\sim 2$~$\mu$s per sample without the morphology computation (so the total time is $\sim 4$~$\mu$s per sample with the morphology computation). Of course, computing the prior restricted to a given morphology will be more expensive, particularly in the librating cases, since one has to reject many more samples. For instance, for the prior on masses that we use in the $75M_\odot$ cases, it takes $\sim 30$ times longer to sample a given number of prior samples from the L$\pi$ morphology than it does to obtain the same number of prior samples with no restriction on the morphology, while it just takes $\sim 10\%$ longer to obtain the same number of C morphology samples. These times (and the time for the L$0$ morphology samples) are in line with the fractions of samples in the different morphologies of $91\%$, $6\%$, and $3\%$ that we find for the prior for the $75M_\odot$ cases.

We find that some of our stochastic sampling runs output parameters sufficiently close to the boundaries between morphologies (the spin-precession resonances) where $S^2_+ = S^2_-$, that very small differences in the parameters (e.g., a fractional change of $10^{-14}$ in one of the masses) can lead to a change in morphology. We noticed this because in a few of these runs, computing the morphology of the nested samples or posterior samples using the output from Bilby gives a morphology other than the one selected in the run, though this is only the case for at most $\sim 0.3\%$ of the total samples.


\section{Statistical methods and test setup}
\label{sec:methods}
In this section we first describe the method we use to constrain the spin morphology of a given BBH and then discuss the simulated GW observations we consider to demonstrate the performance of this method.  

To identify whether a given BBH signal strongly favors a particular morphology, we perform a Bayesian model selection analysis where we compare the statistical evidence for pairs of morphologies. The Bayesian inference framework is routinely employed to infer the properties of detected binary systems (see, e.g.,~\cite{Veitch:2014wba,LIGOScientific:2021djp}) and to perform model selection studies (see, e.g.,~\cite{Veitch:2008ur,Talbot:2017yur,LIGOScientific:2020stg}).
In this framework, the Bayesian evidence ($Z$) quantifies how much the data $d$ prefer a hypothesis (or a model) $H$, and is defined as
	\begin{equation}
	\label{eq-evidence} 
	Z = P(d \vert H) = \int P(\vec{\theta} \vert H ) \, P(d \vert \vec{\theta}, H) d\vec{\theta}, 
	\end{equation} 
where the integrand is the product of the likelihood $P(d \vert \vec{\theta}, H)$ and the prior $P(\vec{\theta} \vert H )$.
Here $\vec{\theta}$ represents the set of binary parameters such as masses and spins. 
Once we choose the appropriate prior boundaries for each model parameter, the integration is done over the entire range of parameters. Here we compute the integral using nested sampling~\cite{Skilling}, as implemented in dynesty~\cite{Speagle2020,dynesty}.
If we have two models $H_{1}$ and $H_{2}$, then the corresponding evidences $Z_{1}$ and $Z_{2}$ can be used to compare the models against data $d$. This comparison is performed using the Bayes factor, defined as
	\begin{equation}
	\label{eq-bf} 
	{\rm BF}_{2}^{1}=\frac{Z_1}{Z_2}. 
	\end{equation} 
A large positive (negative) value of $\log_{10} {\rm BF}^{1}_{2}$ means that model $H_1$ ($H_2$) is preferred over model $H_2$ ($H_1$). Here a standard interpretation of the size of $|\log_{10} {\rm BF}^{1}_{2}|$ (slightly modified from~\cite{Kass:1995loi}) is that values in $[0, 0.5)$ are barely worth mentioning, while values in $[0.5, 1.5)$, $[1.5, 2)$, and $[2, \infty)$, provide positive, strong, and very strong evidence, respectively, in favor of one model over the other. However, to obtain a more quantitative interpretation for a given problem, one can compute a frequentist background of Bayes factors due to noise, as in, e.g.,~\cite{Lo:2018sep}. We leave such investigations for future work.
Hence, by computing the Bayes factor for two spin morphology hypotheses, we can estimate how strongly one morphology is preferred over another for a given BBH.
For example, ${\rm BF}^{\rm L0}_{\rm C}$ and ${\rm BF}^{\rm L\pi}_{\rm C}$ quantify the preference for the L0 and L$\pi$ morphology over the C morphology, respectively, where $Z_{\rm C}$, $Z_{\rm L0}$, and $Z_{\rm L\pi}$ are the evidences computed using Eq.~(\ref{eq-evidence}) with different prior distributions $P(\vec{\theta} \vert H )$. Specifically, for each of these, we restrict the prior to a given morphology using the method given in the previous section, where Bilby automatically normalizes the prior to the constrained region using the ratio of accepted to total samples.

We also compare this full nested sampling Bayes factor calculation with an approximation using the importance weights from a run without restricting the prior to a specific morphology. Specifically, the sum of the weights corresponding to the nested samples in a given morphology gives an approximation to the fractional evidence in that morphology multiplied by the fractional prior volume in that morphology. This sort of approximation was suggested by Skilling~\cite{skilling2004}. For example, denoting two morphologies by I and II, we have
\<
{\rm BF}^\text{II}_\text{I} \simeq \frac{\left(\sum_{M(\vec{\theta}^\text{NS}_k) \,=\, \text{II}}w_k\right)\left(\int_\text{I}P(\vec{\theta} \vert H )d\vec{\theta}\right)}{\left(\sum_{M(\vec{\theta}^\text{NS}_k) \,=\, \text{I}}w_k\right)\left(\int_\text{II}P(\vec{\theta} \vert H )d\vec{\theta}\right)},
\?
where $\{\vec{\theta}^\text{NS}_k\}$ denotes the set of nested samples, with importance weights $\{w_k\}$, and $M(\vec{\theta}^\text{NS}_k)$ returns the morphology of the binary with parameters $\vec{\theta}^\text{NS}_k$. The integrals restrict to the parameters that give the morphology indicated. We approximate these integrals using a Monte Carlo sum, drawing $10^6$ points from the prior distribution. We estimate the errors in this approximation using the method suggested by Skilling~\cite{Skilling}, where one computes the distribution of evidences obtained when resampling the weights. Here we resample the weights $100$ times and compute the standard deviation of the distribution of the log evidence, for uniformity with the use of a standard deviation of the log evidence in the dynesty error estimate. We find that applying this method to the full set of nested samples (with no restriction on the morphology) produces an error on the log evidence that is $\sim 30$--$40\%$ smaller than the dynesty error estimate. However, these error estimates are all small enough that they are negligible for our purposes.

We occasionally also use the simple method of counting the number of posterior samples in each morphology, where the ratio between the numbers of samples in two morphologies gives a crude approximation to the Bayes factor between the two morphologies, though we find that method actually gives results that agree well with the importance weights approximation in the cases where there are a nonzero number of posterior samples in the morphology. We do this to compare with the use of this method in~\cite{Gangardt:2022ltd}, and also to allow us to assess the effect of the distance prior we use: We use distance marginalization (as described in~\cite{Thrane:2018qnx}), so the distance samples are only reconstructed in postprocessing. The posterior samples are not all statistically independent, since dynesty resamples the nested samples, but we do not concern ourselves with this for the purposes of these checks. In particular, for the comparison with~\cite{Gangardt:2022ltd} we need to use exactly the same procedure they do. We compute a simple error estimate for the Bayes factor in this approximation by estimating the error in the number of samples in a given morphology by $\sqrt{n}$, where $n$ is the number of samples in that morphology. We also checked that the error in the log Bayes factors we obtain with this method agrees to the number of decimal places we quote with the error estimate obtained by taking the standard deviation of the ensemble of log Bayes factors computed using $100$ random draws of half the total number of samples.

To assess the performance of the method discussed above, we simulate a set of mock GW signals from BBHs in different morphologies. 
We use the IMRPhenomXPHM model~\cite{Pratten:2020ceb} for both the simulated GW signals and as the recovery template in our Bayesian inference. 
We choose the components of our binaries to be of comparable mass with mass-ratio $q=m_2/m_1=1/1.2$. This choice is inspired by the fact that comparable mass binaries spend more time in the
L0 or L$\pi$ morphology during their evolution \cite{Schnittman:2004vq}. We choose two redshifted total masses, $20M_\odot$ and $75M_\odot$, for our binaries. These roughly correspond to the first two peaks (at $\sim 8M_\odot$ and $\sim 30M_\odot$) of the observed chirp mass distribution in GWTC-3 \cite{LIGOScientific:2021djp}. Further, we choose three values for the spin magnitudes for the black holes: $\chi_1=\chi_2=0.25, 0.75, 0.95$, as representatives of moderately-, highly-, and close-to-extremally-spinning black holes. 
Finally, we choose two network SNRs, $20$ and $89$, to capture low- and high-SNR observing scenarios. These specific values come from considering the $22M_\odot$ binary with zero spins and the extrinsic parameters given in the caption of Table~\ref{table:angle_pars} at luminosity distances of $1$~Gpc and $250$~Mpc, respectively. Here we choose $250$~Mpc since it is in the $90\%$ credible interval (though on the smaller side) of the luminosity distance for GW170608~\cite{LIGOScientific:2017vox,LIGOScientific:2021usb}, which has a redshifted total mass consistent with $20M_\odot$. We choose $1$~Gpc since many of the events detected in O3 are at about this distance~\cite{LIGOScientific:2020ibl,LIGOScientific:2021usb,LIGOScientific:2021djp}.

\begin{figure*}
\includegraphics[width=0.9\textwidth]{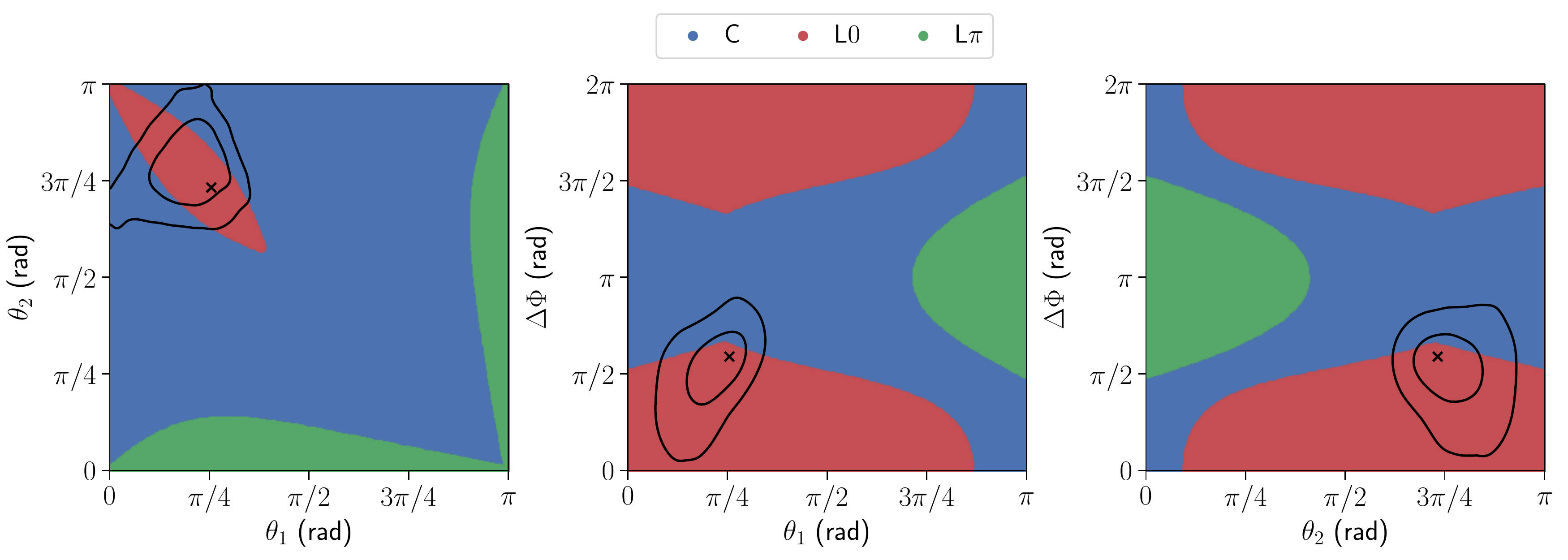}
\caption{\label{fig:morph_par_dep} The spin angle dependence of the morphology for the $75M_\odot$, $\chi_1 = \chi_2 = 0.75$, L$0$ boundary case. The contours show the $50\%$ and $90\%$ credible level for the spin angles from the standard IMRPhenomXPHM analysis (with no cut on morphology) of the SNR $89$ case. The L$0$ boundary parameters ($\theta_1 = 0.8$, $\theta_2 = 2.3$, $\Delta\Phi = 1.8$, all in radians) are given by the cross, and each of the 2d plots is a cut through the 3d spin angle parameter space at one of these parameters.}
\end{figure*}

We find that it is easier to distinguish between two morphologies when the binary's parameters are far from the boundaries between morphologies, particularly in the spin angle subspace---we illustrate the dependence of the morphology on the spin angles in Fig.~\ref{fig:morph_par_dep}. For each total mass and spin magnitude in our simulated observations, we choose spin angles $\theta_1, \theta_2$, and $\Delta \Phi$ such that a given binary is well within each morphology (central cases) as well as close to the boundary of each morphology (boundary cases). For the central cases we have, (i) C central, (ii) L0 central, and (iii) L$\pi$ central, while for the boundary cases we have (iv) C near L0 boundary, (v) C near L$\pi$ boundary, (vi) L0 boundary (with C), and (vii) L$\pi$ boundary (with C). 
The spin angles corresponding to these cases are given in Table~\ref{table:angle_pars} along with the distances we use to obtain an SNR of $89$. We find that L0 boundary and L$\pi$ boundary cases have transitioned to L0 and L$\pi$ morphology only shortly (sometimes by less than a precessional cycle) before 
the binary reached the reference frequency (here $20$~Hz), while those far from the boundary have been in the librating morphology for longer (more than $10$ precessional cycles, in some cases).

\begin{table*}
\caption{\label{table:angle_pars} The spin angles (in radians) at the reference frequency of $20$~Hz used for the different morphology cases we consider and the luminosity distances that give a network SNR of $89$ with IMRPhenomXPHM and the extrinsic parameters we use: $\text{right ascension} = 0.5$~rad,  $\text{declination} = -0.5$~rad, $\text{inclination angle}=\pi/3$~rad, $\text{polarization angle}=0$ rad, and a GPS time of arrival at the geocenter of $1180922494.5$. The spin angles and distances are given for the two redshifted total masses we consider ($20M_\odot$ and $75M_\odot$). To obtain the distances for the SNR $22$ cases, multiply each distance by $89/22 \simeq 4$. All binaries have the same mass ratio of $1/1.2$.}
\begin{tabular}{*{25}{c}}
\hline\hline
& & \multicolumn{7}{c}{$\chi_1 = \chi_2 = 0.25$} & & \multicolumn{7}{c}{$\chi_1 = \chi_2 = 0.75$} & & \multicolumn{7}{c}{$\chi_1 = \chi_2 = 0.95$}\\
\cline{3-9}
\cline{11-17}
\cline{19-25}
Type & & $\theta_1$ & & $\theta_2$ & & $\Delta\Phi$ & & $d_L$ [Mpc] & & $\theta_1$ & & $\theta_2$ & & $\Delta\Phi$ & & $d_L$ [Mpc] & & $\theta_1$ & & $\theta_2$ & & $\Delta\Phi$ & & $d_L$ [Mpc]\\
\hline
\multicolumn{25}{c}{$20M_\odot$}\\
\hline
C central & \qquad\qquad\qquad & $1.6$ & \quad & $1.6$ & \quad & $3.0$ & \quad & $256$ & \qquad\qquad\qquad & $1.2$ & \quad & $2.2$ & \quad & $3.0$ & \quad & $252$ & \qquad\qquad\qquad & $1.3$ & \quad & $2.5$ & \quad & $3.0$ & \quad & $250$\\
C near L$0$ boundary & \qquad\qquad\qquad & $1.2$ & \quad & $2.6$ & \quad & $1.1$ & \quad & $258$ & \qquad\qquad\qquad & $1.2$ & \quad & $2.2$ & \quad & $1.9$ & \quad & $267$ & \qquad\qquad\qquad & $1.3$ & \quad & $2.5$ & \quad & $1.9$ & \quad & $265$\\
C near L$\pi$ boundary & \qquad\qquad\qquad & $2.8$ & \quad & $2.6$ & \quad & $3.0$ & \quad & $246$ & \qquad\qquad\qquad & $2.2$ & \quad & $2.2$ & \quad & $3.0$ & \quad & $244$ & \qquad\qquad\qquad & $2.2$ & \quad & $2.5$ & \quad & $3.0$ & \quad & $238$\\
L$0$ central & \qquad\qquad\qquad & $1.4$ & \quad & $2.9$ & \quad & $0.3$ & \quad & $255$ & \qquad\qquad\qquad & $0.8$ & \quad & $2.3$ & \quad & $0.6$ & \quad & $273$ & \qquad\qquad\qquad & $0.8$ & \quad & $2.3$ & \quad & $0.7$ & \quad & $275$\\
L$0$ boundary & \qquad\qquad\qquad & $1.4$ & \quad & $2.9$ & \quad & $1.1$ & \quad & $256$ & \qquad\qquad\qquad & $0.8$ & \quad & $2.3$ & \quad & $1.6$ & \quad & $276$ & \qquad\qquad\qquad & $0.8$ & \quad & $2.3$ & \quad & $1.8$ & \quad & $288$\\
L$\pi$ central & \qquad\qquad\qquad & $1.3$ & \quad & $0.1$ & \quad & $3.2$ & \quad & $259$ & \qquad\qquad\qquad & $1.2$ & \quad & $0.3$ & \quad & $3.2$ & \quad & $269$ & \qquad\qquad\qquad & $1.3$ & \quad & $0.3$ & \quad & $3.2$ & \quad & $271$\\
L$\pi$ boundary & \qquad\qquad\qquad & $1.3$ & \quad & $0.1$ & \quad & $2.0$ & \quad & $260$ & \qquad\qquad\qquad & $1.2$ & \quad & $0.3$ & \quad & $2.0$ & \quad & $274$ & \qquad\qquad\qquad & $1.3$ & \quad & $0.3$ & \quad & $1.9$ & \quad & $278$\\
\hline
\multicolumn{25}{c}{$75M_\odot$}\\
\hline
C central & \qquad\qquad\qquad & $1.6$ & \quad & $1.6$ & \quad & $3.0$ & \quad & $720$ & \qquad\qquad\qquad & $1.3$ & \quad & $2.5$ & \quad & $3.0$ & \quad & $808$ & \qquad\qquad\qquad & $1.3$ & \quad & $2.5$ & \quad & $3.0$ & \quad & $809$\\
C near L$0$ boundary & \qquad\qquad\qquad & $1.2$ & \quad & $2.2$ & \quad & $1.1$ & \quad & $753$ & \qquad\qquad\qquad & $1.3$ & \quad & $2.5$ & \quad & $2.0$ & \quad & $861$ & \qquad\qquad\qquad & $1.3$ & \quad & $2.5$ & \quad & $2.0$ & \quad & $892$\\
C near L$\pi$ boundary & \qquad\qquad\qquad & $2.5$ & \quad & $2.2$ & \quad & $3.0$ & \quad & $675$ & \qquad\qquad\qquad & $2.1$ & \quad & $2.5$ & \quad & $3.0$ & \quad & $608$ & \qquad\qquad\qquad & $2.0$ & \quad & $2.5$ & \quad & $3.0$ & \quad & $529$\\
L$0$ central & \qquad\qquad\qquad & $0.3$ & \quad & $2.5$ & \quad & $0.3$ & \quad & $718$ & \qquad\qquad\qquad & $0.8$ & \quad & $2.3$ & \quad & $0.9$ & \quad & $738$ & \qquad\qquad\qquad & $0.8$ & \quad & $2.3$ & \quad & $0.9$ & \quad & $724$\\
L$0$ boundary & \qquad\qquad\qquad & $0.3$ & \quad & $2.5$ & \quad & $1.1$ & \quad & $709$ & \qquad\qquad\qquad & $0.8$ & \quad & $2.3$ & \quad & $1.9$ & \quad & $723$ & \qquad\qquad\qquad & $0.8$ & \quad & $2.3$ & \quad & $1.9$ & \quad & $739$\\
L$\pi$ central & \qquad\qquad\qquad & $1.4$ & \quad & $0.2$ & \quad & $3.1$ & \quad & $733$ & \qquad\qquad\qquad & $1.1$ & \quad & $0.4$ & \quad & $3.2$ & \quad & $792$ & \qquad\qquad\qquad & $1.0$ & \quad & $0.5$ & \quad & $3.2$ & \quad & $828$\\
L$\pi$ boundary & \qquad\qquad\qquad & $1.4$ & \quad & $0.2$ & \quad & $2.1$ & \quad & $737$ & \qquad\qquad\qquad & $1.1$ & \quad & $0.4$ & \quad & $2.0$ & \quad & $796$ & \qquad\qquad\qquad & $1.1$ & \quad & $0.5$ & \quad & $2.0$ & \quad & $815$\\
\hline\hline
\end{tabular}
\end{table*}

For each simulated observation, we choose the luminosity distance to ensure that the network SNR is fixed at $22$ or $89$ in a three-detector network consisting of the two LIGO detectors and the Virgo detector operating with their plus-era (O5) sensitivities~\cite{Aasi:2013wya}, using the more sensitive Virgo noise curve.\footnote{We do not include the KAGRA detector, since the latest predictions for its sensitivity in O5~\cite{timeline_graphic} are uncertain by almost an order of magnitude.} The other extrinsic parameters are chosen to be the same for 
all simulated observations and are given in the caption of Table~\ref{table:angle_pars}. For all cases, the likelihood function is computed with the lower cut-off frequency fixed to $f_\text{low}=20$~Hz and the upper cut-off frequency is given by $f_\text{high}=\alpha^\text{roll-off} \times \{\text{Nyquist frequency}\}$, where $\alpha^\text{roll-off}=0.875$ is introduced to account for the effects of a window function (as discussed in Appendix~E of~\cite{LIGOScientific:2021djp}) and the Nyquist frequency is $1024$~Hz for the $20M_\odot$ cases and $512$~Hz for the $75M_\odot$ cases. We do not include noise in the simulated observations to avoid parameter biases (i.e., we use the zero realization of the noise).

We use the same agnostic priors used in the standard LIGO-Virgo-KAGRA (LVK) analyses (discussed in, e.g., Appendix~E~3 of~\cite{LIGOScientific:2021djp}), notably priors that are uniform in the redshifted component masses and spin magnitudes and isotropic in spin directions,
binary orientation, and sky location. Besides the additional restriction to a given morphology to compute the Bayes factors, the only difference is that we use a prior on the luminosity distance $d_L$ that is uniform in Euclidean volume (so $\propto d_L^2$), as in earlier LIGO-Virgo analyses (e.g., \cite{LIGOScientific:2018mvr}), as opposed to the more complicated prior uniform in the comoving frame merger rate that is used in the latest LVK analyses~\cite{LIGOScientific:2021djp}. While we do find that there are correlations between the luminosity distance and the mass ratio and spins in some cases, we have checked that reweighting to the same prior that the LVK uses for its latest results makes a negligible difference in our results, at most a change of $0.1$ in the estimated $\log_{10}$ Bayes factors from the sample counting calculation.


\section{Results}
\label{sec:results}
\subsection{Results for simulated observations}

\begin{figure*}[htb]
\includegraphics[width=0.98\textwidth]{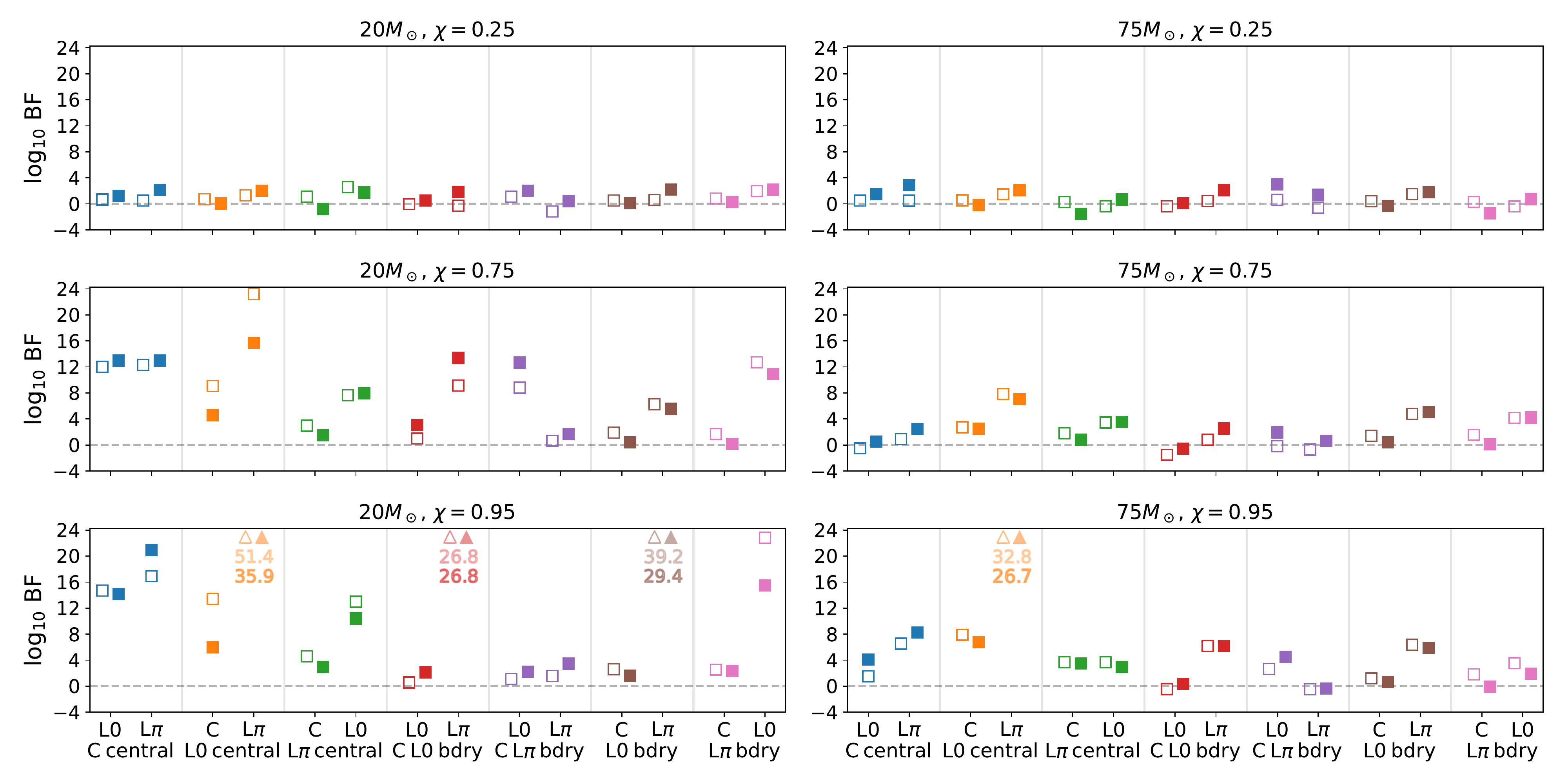}
\caption{\label{fig:BF_high_SNR} The $\log_{10}$ Bayes factors in favor of the true morphology compared to the two alternative morphologies for SNR $89$ BBHs. We give the true morphology on the bottom of the horizontal axis (abbreviating ``boundary'' to ``bdry'') and the two alternative morphologies above it. For an explicit example, the leftmost Bayes factor plotted is $\log_{10} \text{BF}^\text{C}_{\text{L}0}$. We show the results from the nested sampling calculation with filled markers and the importance weights method with unfilled markers. We also use triangles for the cases where both calculations give much larger Bayes factors than the other cases, so the values lie off of the plotted regions, and provide their $\log_{10}$ Bayes factor values. All the nested sampling $\log_{10}$ Bayes factors have errors of $\pm 0.2$, while the importance weights results have errors of at most $\pm 0.6$, but none of these are visible on the scale of this plot.
}
\end{figure*}

We use parallel Bilby~\cite{Ashton:2018jfp,Smith:2019ucc} to compute the Bayes factors of the alternate morphologies with respect to the true morphology and compare them with the importance weights approximation applied to the standard parameter estimation run (with no restriction on the morphology). This thus requires four parameter estimation runs: One restricting to each morphology and one standard run with no restriction on morphology. We plot the results for the SNR $89$ cases in Fig.~\ref{fig:BF_high_SNR}. As expected, we find that the true morphology is most strongly favored compared to both alternative morphologies for highly spinning BBHs with magnitudes $0.75$ and $0.95$. This is because the spin angles, particularly $\Delta\Phi$, are constrained more accurately in these cases. Similarly (and also as expected), the morphologies are in general more strongly constrained for the lower-mass BBHs, where there are more cycles in band, thus leading to stronger constraints on spin angles. Moreover, the true morphology is favored very strongly compared with at least one alternative morphology, particularly for the $0.75$ and $0.95$ spin cases, with $\log_{10}$ Bayes factors in favor of the true morphology larger than $20$ for some cases in the $0.95$ spin case, and often greater than $4$. There is not a significant difference between the Bayes factors for the two total masses in the $0.25$ spin case, where the spin angles are not well constrained. 

We find that the central cases often have the largest Bayes factors in favor of the true morphology, but not always. For instance, for the $0.75$ and $0.95$ spin $20M_\odot$ BBHs, the L$\pi$ boundary case gives a $\text{BF}^{\text{L}\pi}_{\text{L}0}$ that is larger than either of the Bayes factors in the L$\pi$ central case. This is somewhat counterintuitive, since the true value of $\Delta\Phi \simeq 2$~rad in the L$\pi$ boundary cases is closer to the L$0$ morphology region around $\Delta\Phi = 0$. However, in these cases the spin angle posterior falls off quite quickly in the direction of the L$0$ morphology region, about as quickly as it does in the L$\pi$ central cases, even though these have a true value of $\Delta\Phi = 3.2$~rad. This, combined with the fact that the spin angle posterior in the L$\pi$ central cases extends further towards the portion of the L$0$ morphology region close to $\Delta\Phi = 2\pi$~rad, may explain the Bayes factors we obtain.

We also find that the importance weights approximation gives a reasonable approximation to the full nested sampling results, though it tends to overestimate large Bayes factors. The larger differences we find are statistically very significant. For instance, in the $20M_\odot$, $\chi = 0.95$, L$0$ central case, the differences in $\log_{10}\text{BF}^{\text{L}0}_\text{C}$ and $\log_{10}\text{BF}^{\text{L}0}_{\text{L}\pi}$ are $7.2$ and $15.5$, respectively, while the errors in the nested sampling calculation are both $\pm0.2$ and those in the importance weights calculations are $\pm 0.5$ and $\pm 0.6$, respectively. Thus, recalling that the error in the difference is the quadrature sum of the individual errors, the differences correspond to $\sim13\sigma$ and $\sim25\sigma$, respectively. However, there are still a significant number of nested samples in the C and L$\pi$ morphologies in this case: $\sim 3 \times 10^4$ and $\sim 2\times 10^3$, respectively, or $\sim 34\%$ and $\sim 2\%$ of $\sim 9\times 10^4$ total nested samples.

\begin{figure*}[htb]
\includegraphics[width=0.98\textwidth]{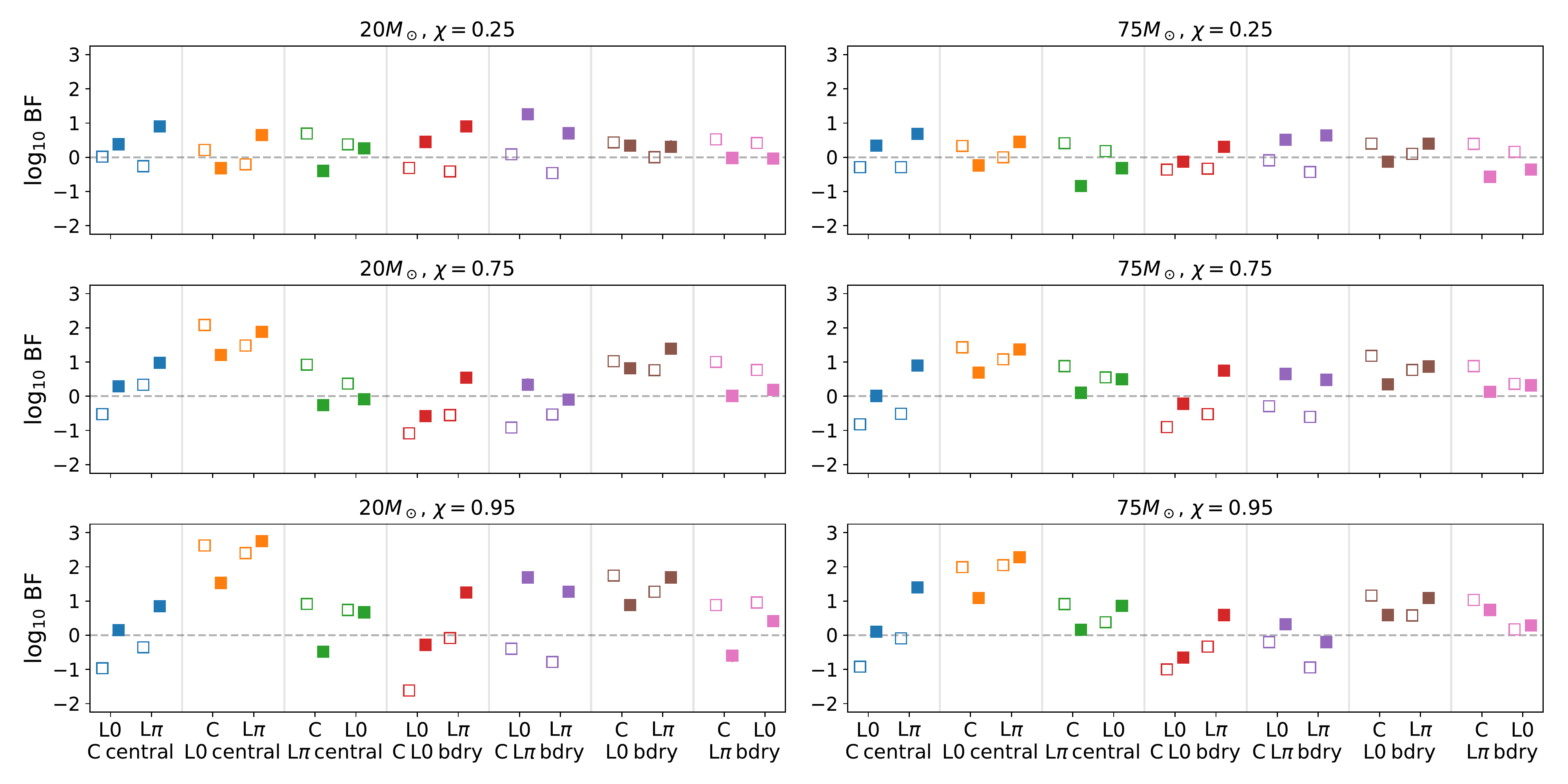}
\caption{\label{fig:BF_low_SNR} The $\log_{10}$ Bayes factors in favor of the true morphology compared to the two alternative morphologies for SNR $22$ BBHs, plotted the same way as in Fig.~\ref{fig:BF_high_SNR} (though note the difference in vertical axis ranges when comparing the two figures). As in Fig.~\ref{fig:BF_high_SNR}, the nested sampling $\log_{10}$ Bayes factors have errors of $\pm 0.2$, which are not visible on the scale of this plot. The importance weights results have errors of $\pm 0.1$, which are not visible either.
}
\end{figure*}

Additionally, for comparison with the relatively large SNR used to obtain the previous results, we also consider a more moderate SNR of $22$, giving the Bayes factors shown in Fig.~\ref{fig:BF_low_SNR}. We find that the true morphology is still favored over both alternative morphologies with a $\log_{10}$ Bayes factor $> 1.2$ in the two higher spin $20M_\odot$ L$0$ central cases as well as the $\chi = 0.95$ $20M_\odot$ C near L$\pi$ boundary case. Additionally, we find that the importance weights calculation is still a reasonable approximation to the full nested sampling results in most cases, as we found in the high-SNR case. However, it gives notably different results than the full nested sampling computation in some cases, notably the $\chi = 0.95$ $20M_\odot$ C near L$\pi$ boundary case, though we have not been able to determine why there is such a significant difference in that particular case. In this case, we also find statistically significant differences of $\sim 8\sigma$ for both $\log_{10}\text{BF}^\text{C}_{\text{L}0}$ and $\log_{10}\text{BF}^\text{C}_{\text{L}\pi}$ and roughly $10\%$ of the $\sim 6\times 10^4$ nested samples are in each of the librating morphologies.

In some low-SNR and/or low-spin cases, we find that the Bayes factors in librating cases still favor the C morphology, notably in the $\chi = 0.25$ $75M_\odot$ high-SNR L$\pi$ central case ($\log_{10}\text{BF}_\text{C}^{\text{L}\pi} = -1.5 \pm 0.2$). In these cases, the spin angles are not well constrained, so the posteriors often fill their prior ranges and thus extend well into the C morphology region, giving significant support for that morphology.

\subsection{Systematic uncertainties}

We make initial investigations of the effects of waveform systematics on our results by computing the Bayes factors using the NRSur7dq4 waveform model~\cite{Varma:2019csw} for injections and recovery in some of the high-mass, high-SNR cases. Future work will consider the extent to which waveform systematics can bias this measurement, by, e.g., analyzing a signal generated with a more accurate waveform model using a faster but less accurate one. Specifically, we chose the cases with the largest and smallest Bayes factor with IMRPhenomXPHM for each choice of spin magnitudes. These cases are: C near L$\pi$ boundary and L$\pi$ central for $0.25$ spin; L$0$ central and C near L$0$ boundary for $0.75$ spin; and L$0$ central and C near L$\pi$ boundary for $0.95$ spin. For each of these, we adjust the distance so that we still obtain an SNR of $89$ with NRSur7dq4 (leading to differences in the distance of $\lesssim 10\%$). 

We use the NRSur7dq4 model since it is constructed in a significantly different way than IMRPhenomXPHM, by directly interpolating waveforms from numerical relativity simulations, and contains more physics, notably all spherical harmonic modes with $\ell \leq 4$ in the coprecessing frame, including the asymmetry between the $\pm m$ modes (in the coprecessing frame) for precessing systems. Recall that IMRPhenomXPHM only contains the $(2, \pm 2)$, $(2, \pm 1)$, $(3, \pm 3)$, $(3, \pm 2)$, and $(4, \pm 4)$ modes in the coprecessing frame, and its $\pm m$ modes in the coprecessing frame are related to each other using the nonprecessing relation. We only made the NRSur7dq4 check for the high-mass case since the model has a limited length, so it is only able to model the dominant mode signal from $20$~Hz for sufficiently high total masses. In fact, NRSur7dq4 waveforms are not long enough for the $|m| = 3, 4$ modes to start at $20$~Hz for a total mass of $75 M_\odot$; they instead start around $21$ and $28$~Hz, respectively, with the exact value depending on the spins. However, we checked that the SNR in the missing portions of the inspiral in the $|m| = 3, 4$ modes is negligible. Note that we are using NRSur7dq4 outside of its training region of spin magnitudes $\leq 0.8$ in the $0.75$ and $0.95$ spin cases. (Significant portions of the spin magnitude posteriors extend above $0.8$ in the $0.75$ spin cases.) However, NRSur7dq4 is found to extrapolate well to higher spin magnitudes. 

We find that in most cases the NRSur7dq4 $\log_{10}$ Bayes factors are greater than the IMRPhenomXPHM ones by between $0.6$ and $1.3$. These correspond to fractional differences in the $\log_{10}$ Bayes factors of $\sim 30\%$ to $\sim 230\%$, where the largest differences correspond to cases where the sign of the Bayes factor changes, though in these cases the individual $\log_{10}$ Bayes factors are small enough that they are only at most $\sim 2$ times the estimated error. However, for the $0.75$ spin L$0$ central case, $\log_{10} \text{BF}^{\text{L}0}_\text{C}$ and $\log_{10} \text{BF}^{\text{L}0}_{\text{L}\pi}$ are $2.4$ and $2.1$ larger, respectively (with fractional differences of $\sim 90\%$ and $\sim 30\%$, respectively), due to much better constraints on $\Delta\Phi$, while for the $0.95$ spin L$0$ central case, $\log_{10} \text{BF}^{\text{L}0}_{\text{L}\pi}$ is smaller by $9$, due to a difference in the shape of the posteriors, though it is still quite large ($17.7$), so the fractional difference is only $\sim 30\%$. The only other cases where the NRSur7dq4 analysis gives a smaller Bayes factor are for the $0.25$ spin C near L$\pi$ boundary case, where $\log_{10} \text{BF}^\text{C}_{\text{L}0}$ and $\log_{10} \text{BF}^\text{C}_{\text{L}\pi}$ are $1.0$ and $0.6$ smaller, respectively, and the $0.95$ spin C near L$\pi$ boundary case, where $\log_{10} \text{BF}^\text{C}_{\text{L}0}$ is $0.8$ smaller; these correspond to fractional differences of $\sim 20\%$ to $\sim 40\%$. Ref.~\cite{Biscoveanu:2021nvg} finds cases where IMRPhenomXPHM gives significantly smaller uncertainties in spin quantities than NRSur7dq4 (see their Fig.~18), though the most dramatic examples they find are for larger total masses than we consider, and we still find that NRSur7dq4 generally gives tighter posteriors on spin quantities than IMRPhenomXPHM in our runs, even in the cases where it finds smaller Bayes factors. In general, while we find nonnegligible differences in the Bayes factors due to the difference in waveform model, the only case where there is a really significant difference is the $0.75$ spin L$0$ central case, where NRSur7dq4 gives a value of $\log_{10} \text{BF}^{\text{L}0}_\text{C}$ that is almost double its value with the IMRPhenomXPHM analysis.

Finally, we consider the uncertainties in our results due to using $1.5$PN expressions for the orbital angular momentum when computing the morphology. We find that the differences compared to using the $2$PN or $2.5$PN expressions are generally small, as expected, given the results discussed in Sec.~\ref{sec:morph}. However, in one case, there is a considerable difference between the results with the $1.5$PN and $2$PN or $2.5$PN orbital angular momentum expressions. As expected, there are also considerable differences between the results with the Newtonian and higher-PN orbital angular momentum expressions in a number of cases. However, in all cases these differences only affect the degree to which a given morphology is favored and do not lead to the wrong morphology being favored. We check these differences using the importance weights calculation, where it is simple to change the order of the orbital angular momentum expressions used in the morphology calculation.

The case with the largest difference is the $75M_\odot$ $0.95$ spin L0 central high-SNR case, where one finds that due to a decrease in the L$\pi$ evidence, the estimated $\log_{10}\text{BF}^{\text{L}0}_{\text{L}\pi}$ is $9.5$ larger with the $1$PN, $2$PN, or $2.5$PN orbital angular momentum expressions than with the $0$PN or $1.5$PN expressions, though it is already quite large ($26.7$) in those cases, so this is only a $36\%$ correction. This difference is because the $1.5$PN orbital angular momentum expressions gives a value that is slightly smaller than that obtained with the $1$PN, $2$PN, or $2.5$PN expressions for two nested samples, making those two nested samples be in the L$\pi$ morphology, and thus increasing the evidence substantially. (The Newtonian expression for the orbital angular momentum also gives a smaller value, but that is expected.) In all other cases, the use of the $2$PN or $2.5$PN orbital angular momentum expressions leads to at most a change of $0.06$ in the $\log_{10}$ Bayes factor, with this largest difference arising from a decrease in $Z_{\text{L}\pi}$ in the $75M_\odot$ $0.95$ spin C near L$\pi$ boundary high-SNR case with IMRPhenomXPHM (there is a decrease of $0.04$ with NRSur7dq4). We find that using the Newtonian orbital angular momentum expression leads to more significant differences in a number of cases, from a decrease of $\log_{10} Z_\text{C}$ by $3.1$ in the $20M_\odot$ $0.75$ spin L$0$ central high-SNR case (giving a $68\%$ increase in  $\log_{10}\text{BF}^{\text{L}0}_\text{C}$) to an increase of $\log_{10} Z_{\text{L}0}$ by $3.6$ in the $20M_\odot$ $0.95$ spin L$\pi$ central high-SNR case (giving a $35\%$ reduction in $\log_{10}\text{BF}^{\text{L}\pi}_{\text{L}0}$). The largest difference due to using the $1$PN orbital angular momentum expression instead of the $1.5$PN one is an increase in $\log_{10} Z_{\text{L}\pi}$ by $0.1$ for the $20M_\odot$ $0.95$ spin C near L$\pi$ boundary high-SNR case.

\subsection{Results for GW200129\_065458}

Since we find that one is able to obtain $\log_{10}$ Bayes factors of $\gtrsim 1$ in favor of the true morphology in a number of the SNR $22$ cases, we thought that it was worth applying this analysis to a real GW event. We choose to analyze the BBH signal GW200129\_065458, which has a median SNR of $27$ and evidence for precession~\cite{LIGOScientific:2021djp, Hannam:2021pit, Varma:2022pld}. The evidence for precession has some caveats, since this event required some glitch mitigation~\cite{Davis:2022ird} and the uncertainties in the glitch subtraction are larger than the difference between precessing and nonprecessing models for the signal~\cite{Payne:2022spz}. Nevertheless, we choose GW200129\_065458 as a relatively high SNR system with possible precession. We use the glitch-subtracted data, noise power spectral densities, and detector calibration uncertainty envelopes from the Gravitational-Wave Open Science Center~\cite{KAGRA:2023pio} and perform the analysis using NRSur7dq4, which is more accurate for this signal than the IMRPhenomXPHM and SEOBNRv4PHM~\cite{Ossokine:2020kjp} models used in the analysis in~\cite{LIGOScientific:2021djp}, as discussed in~\cite{Hannam:2021pit}. We marginalize over detector calibration uncertainties as in~\cite{SplineCalMarg-T1400682}.

We do not find that any morphology is favored, with both $\log_{10}$ Bayes factors consistent with $0$. Specifically, we obtain $\log_{10}\text{BF}_\text{C}^{\text{L}0} = 0.0 \pm 0.2$ and $\log_{10}\text{BF}_\text{C}^{\text{L}\pi} = -0.1 \pm 0.2$. The importance weights (IW) calculation gives $\log_{10}\left[\text{BF}_\text{C}^{\text{L}0}\right]_\text{IW} = 0.3 \pm 0.1$ and $\log_{10}\left[\text{BF}_\text{C}^{\text{L}\pi}\right]_\text{IW} = 0.2 \pm 0.1$. We also give the sample counting (SC) results for comparison with the results in Gangardt~\emph{et al.}~\cite{Gangardt:2022ltd}, where we do not correct for the prior volume, since Gangardt~\emph{et al.}\ do not. These give $\log_{10}\left[\text{BF}_\text{C}^{\text{L}0}\right]_\text{SC} = -0.81 \pm 0.01$ and $\log_{10}\left[\text{BF}_\text{C}^{\text{L}\pi}\right]_\text{SC} = -1.24 \pm 0.02$, which are only slightly larger than those from the prior. We find that $88\%$, $9\%$, and $4\%$ of the prior samples are in the C, L$0$, and L$\pi$ morphologies, respectively (the numbers do not sum to $100\%$ due to rounding), which gives $\log_{10}\left[\text{BF}_\text{C}^{\text{L}0}\right]_\text{prior} \simeq -1.0$ and $\log_{10}\left[\text{BF}_\text{C}^{\text{L}\pi}\right]_\text{prior} \simeq -1.3$.

Our analysis finds significantly fewer samples in the librating morphologies ($17\%$) than that in Gangardt~\emph{et al.}~\cite{Gangardt:2022ltd} ($29\%$). However, the waveforms used for the two analyses are different: We use NRSur7dq4, while Gangardt~\emph{et al.}\ use the combined samples obtained using IMRPhenomXPHM and SEOBNRv4PHM from~\cite{LIGOScientific:2021djp}.\footnote{Gangardt~\emph{et al.}\ use the samples reweighted to a prior that is uniform in comoving volume, but using the samples without the reweighting (and thus the same distance prior we use) only changes the fractions by increasing the fraction of L$0$ samples by a percentage point, due to rounding.} The orbital angular momentum calculations used to obtain the morphology are also different: We use the $1.5$PN expression for the orbital angular momentum in terms of the binary's GW frequency, while Gangardt~\emph{et al.}\ evaluate the Newtonian expression for the orbital angular momentum in terms of the binary's orbital separation using the $2$PN (harmonic coordinate) expression for this separation in terms of the binary's frequency from~\cite{Kidder:1995zr}. We find that the difference in waveforms is the larger effect: If we apply our $1.5$PN morphology calculation to the same samples used by Gangardt~\emph{et al.}, we obtain $26\%$ of the samples in librating morphologies ($15\%$ in L$0$ and $11\%$ in L$\pi$). However, the difference in morphology calculations also has a noticeable effect: If we apply the Gangardt~\emph{et al.}\ morphology calculation to our samples, then we obtain $20\%$ of the samples in the librating morphology ($14\%$ in L$0$ and $6\%$ in L$\pi$, compared to $13\%$ and $5\%$, respectively, with our $1.5$PN morphology calculation).\footnote{If we use the $2$PN or $2.5$PN orbital angular momentum expressions, then the number of samples in the L$0$ and L$\pi$ morphologies decrease by $4\%$ and $2\%$, respectively, so the L$0$ fraction decreases by a percentage point, due to rounding, and the L$\pi$ fraction remains unchanged to the accuracy quoted. Additionally, the $1.5$PN percentages do not sum to the 17\% quoted earlier due to rounding.}


\section{Conclusions and future work}
\label{sec:concl}
The dynamics of spin precession in binary black holes can be classified into three distinct morphologies depending on whether $\Delta \Phi$ oscillates about 0 (L$0$), oscillates about $\pi$ (L$\pi$), or circulates (C) through the full range $[-\pi, \pi]$ over a precession cycle. These morphologies encode information about the binary's formation scenarios at much larger separations \cite{Gerosa:2015tea}, and therefore morphological classification via GW observations can inform us about the astrophysical formation mechanism of the binary \cite{Gerosa:2013laa,Gerosa:2018wbw}. In this paper, we developed a Bayesian model selection method to constrain spin morphologies in binary black holes from their GW signals. The code is available as a Python package~\cite{morph_package} which can be used directly to set the prior for a specific morphology with the parameter estimation code Bilby~\cite{Ashton:2018jfp, Romero-Shaw:2020owr}.

We demonstrated the performance of our method on various simulated GW signals of comparable masses ($q=1/1.2$) with different total detector-frame masses ($20M_\odot$, $75M_\odot$), spin magnitudes ($0.25$, $0.75$, $0.95$) and SNRs ($20$, $89$). For each of the simulated binary configurations, we chose spin tilts such that the binary is away from and close to the boundary of a particular morphology in the spin angle parameter space. In particular, we considered seven cases for each total mass, spin magnitude and SNR: C central, C near L$0$ boundary, C near L$\pi$ boundary, L$0$ central, L$0$ boundary (with C), L$\pi$ central, and L$\pi$ boundary (with C). 
We used the IMRPhenomXPHM model~\cite{Pratten:2020ceb} to simulate the signals and as the recovery template in parameter estimation analysis.

We found that one favors the true morphology with $\log_{10}$ Bayes factors $\geq4$ in optimistic cases, e.g., high spins and higher SNRs. This is due to better constrained spin parameters. However, one can exclude at least one morphology with a $\log_{10}$ Bayes factor $\sim2$ for some less optimistic cases with, e.g., smaller spins or lower SNRs. We also found that the importance weights approximation for the Bayes factors gives reasonable agreement with nested sampling in many cases, though it overestimates the Bayes factors in most cases where an alternative morphology is very strongly disfavored. 

We also investigated the effects of waveform systematics on our Bayes factor results using the NRSur7dq4 waveform model~\cite{Varma:2019csw} for injections and recovery. We considered only the high mass ($75M_\odot$) and high-SNR ($89$) cases for each spin magnitude injection. We found that in most cases the NRSur7dq4 $\log_{10}$ Bayes factors are greater than the IMRPhenomXPHM ones by between $0.6$ and $1.3$. However, in one case, the NRSur7dq4 $\log_{10}$ Bayes factors are significantly larger (increases of $\sim 2$) than those from IMRPhenomXPHM due to better constraints on $\Delta \Phi$, while in a few other cases the NRSur7dq4 Bayes factors are smaller than those obtained with IMRPhenomXPHM. We also applied the analysis (using NRSur7dq4) to GW200129\_065458 due to its relatively large SNR ($\sim 27$) and evidence for precession. 
We found that GW200129\_065458 does not favor any morphology with all $\log_{10}$ Bayes factors consistent with zero.

In the future, we will develop methods to constrain the fraction of binaries in a given morphology in the population, following~\cite{Saleem:2021vph}, and subsequently apply these methods to GWTC-3 events. Additionally, we will investigate whether one obtains better constraints on the morphology at a dimensionless reference frequency or time (as suggested in~\cite{Varma:2021csh,Varma:2021xbh}), rather than the dimensionful reference frequency we consider. We will also consider whether inferring the morphology at the reference frequency gives stronger constraints on the binary's tilt angles at formation (e.g., distinguishing between the standard and reversed mass ratio scenarios in binaries with efficient tides, as discussed in~\cite{Gerosa:2013laa}) than one obtains by directly computing the posterior on the tilt angles at formally infinite separation, as in~\cite{Johnson-McDaniel:2021rvv,Kulkarni:2023nes}. In addition to these studies, we will also consider the extent to which waveform systematics can bias the inference of the morphology by analyzing a more accurate waveform (e.g., produced with NRSur7dq4 for the higher masses for which it is applicable or SEOBNRv5PHM~\cite{Ramos-Buades:2023ehm} for lower masses) with a faster but less accurate model, such as the IMRPhenomXPHM model that we used for our primary analysis (and is applicable to a wider parameter space than NRSur7dq4 is). Since SEOBNRv5PHM is only slightly slower than IMRPhenomXPHM for lower masses such as the $20M_\odot$ total mass we consider, we could also use it to analyze real signals in addition to IMRPhenomXPHM as a check of waveform systematics.

\acknowledgments

We thank Daria Gangardt, Davide Gerosa, Leo Stein, and Aditya Vijaykumar for useful discussions. We also thank Sylvia Biscoveanu and the anonymous referees for a careful reading of the paper and useful comments and suggestions. NKJ-M is supported by NSF grant AST-2205920. KSP acknowledges support from the Dutch Research Council (NWO). NVK is thankful to the Max Planck Society's Independent Research Group Grant. AG is supported in part by NSF grants PHY-2308887 and AST-2205920.
The authors are grateful for computational resources provided by the LIGO Lab and supported by NSF Grants PHY-0757058 and PHY-0823459. We also acknowledge the use of the Maple cluster at the University of Mississippi (funded by NSF Grant CHE-1338056), the IUCAA LDG cluster Sarathi, the University of Birmingham's BlueBEAR HPC service, Nikhef's Visar cluster, and Max Planck Computing and Data Facility's clusters Raven and Cobra for the computational/numerical work.

This research has made use of data or software obtained from the Gravitational Wave Open Science Center (gwosc.org)~\cite{KAGRA:2023pio}, a service of LIGO Laboratory, the LIGO Scientific Collaboration, the Virgo Collaboration, and KAGRA. LIGO Laboratory and Advanced LIGO are funded by the United States National Science Foundation (NSF) as well as the Science and Technology Facilities Council (STFC) of the United Kingdom, the Max-Planck-Society (MPS), and the State of Niedersachsen/Germany for support of the construction of Advanced LIGO and construction and operation of the GEO600 detector. Additional support for Advanced LIGO was provided by the Australian Research Council. Virgo is funded, through the European Gravitational Observatory (EGO), by the French Centre National de Recherche Scientifique (CNRS), the Italian Istituto Nazionale di Fisica Nucleare (INFN) and the Dutch Nikhef, with contributions by institutions from Belgium, Germany, Greece, Hungary, Ireland, Japan, Monaco, Poland, Portugal, Spain. KAGRA is supported by Ministry of Education, Culture, Sports, Science and Technology (MEXT), Japan Society for the Promotion of Science (JSPS) in Japan; National Research Foundation (NRF) and Ministry of Science and ICT (MSIT) in Korea; Academia Sinica (AS) and National Science and Technology Council (NSTC) in Taiwan.

This study used the software packages dynesty~\cite{Speagle2020,dynesty}, LALSuite~\cite{LALSuite}, Matplotlib~\cite{Hunter:2007}, NumPy~\cite{harris2020array}, parallel Bilby~\cite{Ashton:2018jfp,Smith:2019ucc}, PESummary~\cite{Hoy:2020vys}, and seaborn~\cite{waskom2020seaborn}. This is LIGO document P2300012.

\bibliography{morph_paper}

\end{document}